\def\year{2021}\relax
\documentclass[letterpaper]{article} 
\usepackage{aaai21}  
\usepackage{times}  
\usepackage{helvet} 
\usepackage{courier}  
\usepackage[hyphens]{url}  
\usepackage{graphicx} 
\usepackage{natbib}  
\usepackage{caption} 
\usepackage{amsmath,amsfonts,bm}

\def\eqref#1{equation~\ref{#1}}

\def\1{\bm{1}}

\def\vv{{\bm{v}}}

\DeclareMathAlphabet{\mathsfit}{\encodingdefault}{\sfdefault}{m}{sl}
\SetMathAlphabet{\mathsfit}{bold}{\encodingdefault}{\sfdefault}{bx}{n}

\DeclareMathOperator*{\argmax}{arg\,max}
\DeclareMathOperator*{\argmin}{arg\,min}

\usepackage{comment}
\usepackage{threeparttable}
\usepackage{bm}
\usepackage{multirow}
\usepackage{amsthm}
\usepackage{color}

\usepackage{subfigure}
\usepackage{wrapfig}

\usepackage{algorithmic}
\usepackage[ruled,linesnumbered]{algorithm2e}
\usepackage{adjustbox}

\usepackage{booktabs}       

\usepackage{mydefs}
\title{Learning to Recommend from Sparse Data via Generative User Feedback}
\author{

    Wenlin Wang \textsuperscript{\rm 1},
    Hongteng Xu \textsuperscript{\rm 2},
    Ruiyi Zhang \textsuperscript{\rm 1},
    Wenqi Wang \textsuperscript{\rm 3},
    Piyush Rai \textsuperscript{\rm 4},
    Lawrence Carin \textsuperscript{\rm 1}
}
\affiliations{

    \textsuperscript{\rm 1} Department of ECE, Duke University, 
    \textsuperscript{\rm 2} Gaoling School of Artificial Intelligence, Renmin University of China,\\
    \textsuperscript{\rm 3} Facebook, Inc,
    \textsuperscript{\rm 4} Department of Computer Science and Engineering, IIT Kanpur\\
    wlwang616@gmail.com

}

\begin{document}

\maketitle

\begin{abstract}
Traditional collaborative filtering (CF) based recommender systems tend to perform poorly when the user-item interactions/ratings are highly scarce. To address this, we propose a learning framework that improves collaborative filtering with a synthetic feedback loop (CF-SFL) to simulate the user feedback. The proposed framework consists of a ``recommender'' and a ``virtual user''. The ``recommender'' is formulated as a CF model, recommending items according to observed user preference. The ``virtual user'' estimates rewards from the recommended items and generates a \emph{feedback} in addition to the observed user preference. The ``recommender'' connected with the ``virtual user'' constructs a closed loop, that recommends users with items and imitates the \emph{unobserved} feedback of the users to the recommended items. The synthetic feedback is used to augment the observed user preference and improve recommendation results. Theoretically, such model design can be interpreted as inverse reinforcement learning, which can be learned effectively via rollout (simulation). Experimental results show that the proposed framework is able to enrich the learning of user preference and boost the performance of existing collaborative filtering methods on multiple datasets.
\end{abstract}

\section{Introduction}
Recommender systems are important modules for abundant online applications, helping users explore items of potential interest. 
As one of the most effective approaches, collaborative filtering~\cite{sarwar2001item,koren2015advances,he2017neural} and its deep neural networks based variants~\cite{he2017neural,wu2016collaborative,liang2018variational,li2017collaborative,yang2017bridging,wang2018neural} have been widely studied. These methods leverage patterns across similar users and items, predicting user preferences and have demonstrated encouraging results in recommendation tasks~\cite{bennett2007netflix,hu2008collaborative,schedl2016lfm}.
Among these works, beside ``user-item'' pair data  (e.g., ratings or interaction/purchase history), side information, $e.g.$, user reviews and scores on items, has also been leveraged~\cite{menon2011response,fang2011matrix}. 
Such side information is a kind of user feedback to the recommended items, which is often useful for improving the recommendation systems.

Unfortunately, the user-item pairs and user feedback are extremely sparse as compared to the search space of items.
Moreover, when the recommendation systems are trained on static observations, the user feedback is unavailable until it is deployed in real-world applications; in both training and validation phases, the target systems have no access to any feedback because no one has observed the recommended items.
Therefore, the recommendation systems may suffer from overfitting, and their performance may degrade accordingly, especially in the initial phase of deployment.
Although real-world recommendation systems are usually updated in an online manner with the help of increasing observed
user preference~\cite{rendle2008online,agarwal2010fast,he2016fast}, introducing a feedback learning mechanism \emph{during} their training phases can potentially improve the system that is eventually deployed. However, this aspect has largely been neglected by the existing recommender system frameworks. 

\begin{figure}
	\centering
	\includegraphics[width=0.43\textwidth]{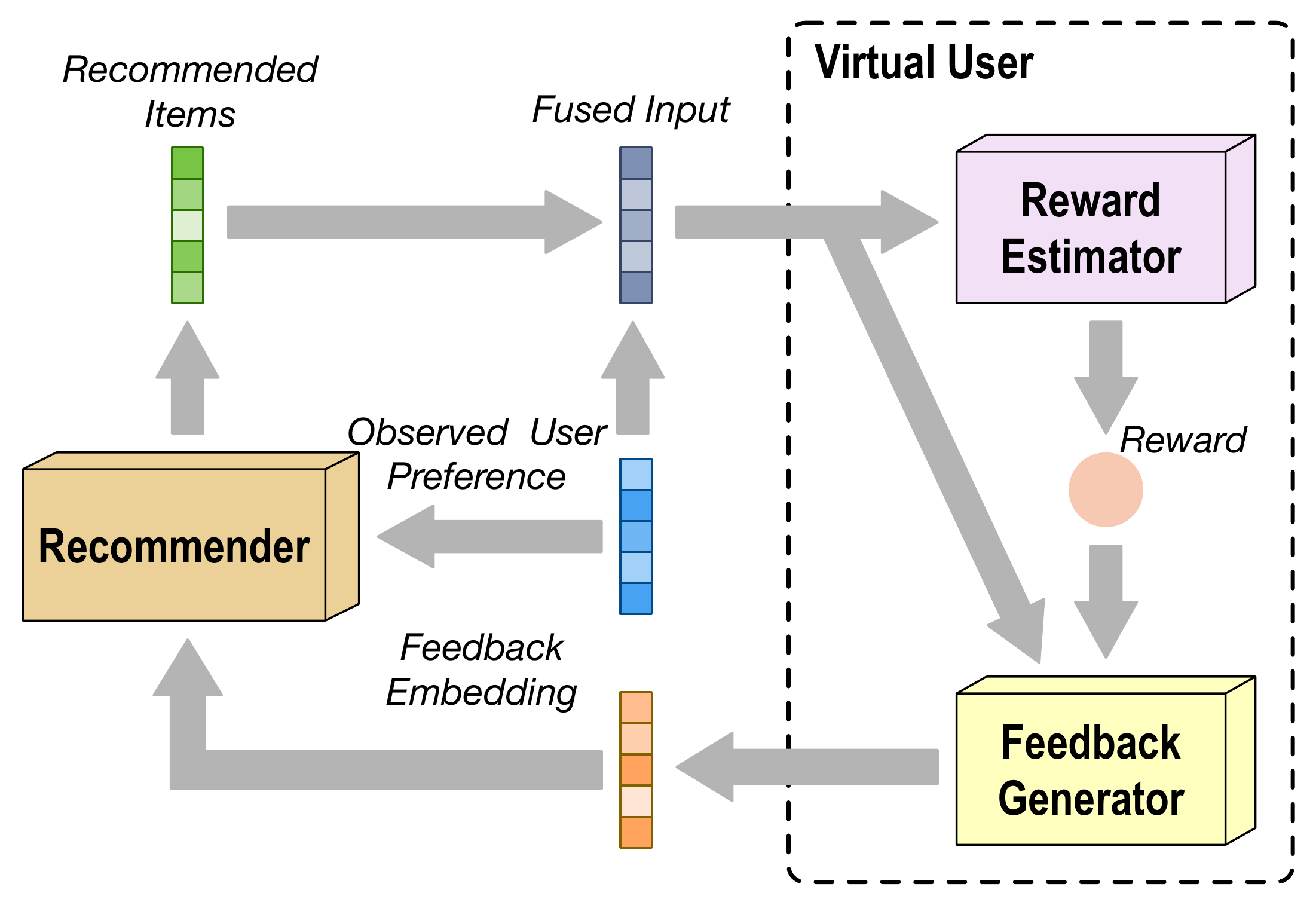}
	\vspace{-1.0em}
	\caption{\small Illustration of our proposed CF-SFL framework for collaborative filtering.}\label{fig:scheme}
\end{figure}

Motivated by these observations, we propose a novel framework that achieves collaborative filtering with a synthetic feedback loop (CF-SFL).
As shown in Figure~\ref{fig:scheme}, the proposed framework consists of a ``recommender'' and a ``virtual user.'' The recommender is a collaborative filtering (CF) model, which predicts items based on observed user preference.
The observed user preference vector reflects intrinsic preferences of the user, while the recommended items vector represents the model's estimated preferences of the user on the items. 
Taking a combination of the observed user preference and the recommended items (the estimated user preference) as inputs, which we refer to as \emph{fused input} (cf., Figure~\ref{fig:scheme}), the virtual user, which is the key aspect of our model, imitates a real-world setting and provides a \emph{synthesized} user feedback (in form of an embedding) to the CF module.
In particular, the virtual user contains a \textit{reward estimator} and a \textit{feedback generator}. The \textit{reward estimator} estimates rewards based on the fused inputs (the combined representation of the user observation and its recommended items), learned with a generative adversarial regularizer. The \textit{feedback generator}, conditioned on the estimated rewards as well as the fused inputs, generates the feedback embedding to augment the original user preference vector. 
Such a framework constructs a closed loop between the target CF model and the virtual user, synthesizing user feedback as side information to improve recommendation results. 

The proposed CF-SFL framework can be interpreted as an inverse reinforcement learning (IRL) set-up, in which the recommender learns to recommend the user items (policy) with the estimated guidance (feedback) from the proposed virtual user. 
The proposed feedback loop can be understood as an effective rollout procedure for recommendation, jointly updating the recommender (policy) and the virtual user (consisting of the reward estimator and the feedback generator).
Essentially, even if side information ($i.e.$, real-world user feedback) is \emph{unavailable}, our model is still applicable to \emph{synthesize} feedback during both training as well as inference phases.
The proposed framework is general and the recommender module can use any of the various existing CF methods, making our framework significantly modular. 
A comprehensive set of experimental results show that the performance of existing CF models can be remarkably improved within the proposed framework.

\section{Proposed Framework}
In this section, we first describe the problem set-up and provide a detailed description of each module that is part of the proposed framework.
\subsection{Problem statement}
Suppose we have $N$ users with $M$ items in total. We denote the observed user-item matrix as $\Xv=[\xv_i] \in \{0, 1\}^{N\times M}$, where each vector $\xv_i=[x_{ij}]\in \mathcal{R}^M$, $i=1,...,N$, represents the \emph{observed} user preferences for the $i$-th user. 
Here $x_{ij} = 1$ indicates the the $j$-th item is bought or clicked by the $i$-th user; otherwise the $j$-th item is either irrelevant to the $i$-th user or we do not have knowledge about their relationship.  
A recommendation system outputs the estimated user preferences, denoted as $\av_i=[a_{ij}]\in\mathcal{R}^{M}$, whose element $a_{ij}$ indicates the estimated preference of the $i$-th user to the $j$-th item. 
The system can then recommend each user the top few items (e.g., given a specified consumption budget) based on the value of the estimated preferences $a_{ij}$'s. 

In practice, for each user, the vector $\xv_i$ just contains partial information about the user's preferences on items and an ideal recommendation system works dynamically with a closed loop --- users often generate feedback on the recommended items while the system considers these feedback to revise recommended items in the future. 
Formally, this \emph{feedback-driven} recommendation process can be written formally as
\begin{eqnarray}\label{eq:recsys}
\begin{aligned}
\av_i^{t} = \pi(\xv_i, \vv_i^{t}),~~ \vv_i^{t+1} = f(\xv_i, \av_i^{t}),~~ \mbox{for}~i=1,...,N,
\end{aligned}
\end{eqnarray}
where $\pi(\cdot)$ represents the target recommender while $f(\cdot)$ represents the coupled feedback mechanism of the system. Here,
$\vv_i\in \mathcal{R}^{d}$ denotes the embedding of the aggregated user feedback on the previously recommended items. 
At time-step $t$, the recommender predicts preferred items according to the observed user preference $\xv_i$ and previous feedback $\vv_i^t$, subsequently, the user generates an updated feedback $\vv_i^{t+1}$ to the recommender. 
Note that Eq.~(\ref{eq:recsys}) is different from existing sequential recommendation models~\cite{mishra2015web,wang2016spore} because these methods do not have the feedback loop, and they just update the recommender module $\pi$ according to observed sequential observations, \textit{i.e.}, $\xv_i^t$ for different time-steps $t$'s.\footnote{When the static observation $\xv_i$ in Eq.~(\ref{eq:recsys}) is replaced with sequential observation $\xv_i^t$, Eq.~(\ref{eq:recsys}) is naturally extended to a sequential recommendation system with a feedback loop. In this work, we focus on the case with static observations and train a recommender system accordingly.}

Unfortunately, the feedback information is often unavailable during training and inference.
Accordingly, most existing collaborative filtering-based recommendation methods do not have a feedback loop in the system, and learn the recommendation system purely from the statically observed user-item matrix $\Xv$~\cite{liang2018variational,li2017collaborative}. 
Although, in some settings, side information like user reviews is associated with the observation matrix, the methods using such information often treat it as a source of static knowledge rather than a dynamic feedback. 
They mainly focus on fitting the ground-truth recommended items with the recommender $\pi(\cdot)$ given fixed $\xv_i$'s and fixed $\vv_i$'s, while ignoring the whole recommendation-feedback loop in Eq.~(\ref{eq:recsys}).
Without a feedback mechanism, $f(\cdot)$, $\pi(\cdot)$ may overfit the (possibly scarce) user observations and the static side information, especially in dynamic settings.

To overcome the aforementioned problems, we propose a collaborative filtering framework with a synthetic feedback loop (CF-SFL). As shown in Figure~\ref{fig:scheme}, besides the traditional recommendation module, the proposed framework further introduces a \emph{virtual user}, which imitates the recommendation-feedback loop, even if the real user feedback is unavailable.

\subsection{The recommender}
In our framework, the recommender implements the function $\pi(\cdot)$ in Eq.~(\ref{eq:recsys}), which takes the observed user preference $\xv_i$ and the user's previous feedback embedding $\vv_i^{t}$ as inputs and recommends items accordingly. 
In principle, the recommender $\pi(\cdot)$ can be defined with high flexibility, which can be based on any existing shallow/deep collaborative filtering method that predicts items from user representations, such as
WMF~\cite{hu2008collaborative}, CDAE~\cite{wu2016collaborative},
VAE~\cite{liang2018variational}, etc.

Our goal is to mimic and integrate dynamic user feedback and therefore, in this work, we develop an inverse reinforcement learning (IRL) based framework, adapted for CF. To the best of our knowledge, none of the existing CF approaches incorporate such a dynamic feedback loop, and our work is the first such attempt in this direction.

In particular, the recommendation-feedback loop generates a sequence of interactions between each user and the recommender, $i.e.$, $(\sv_i^t, \av_i^t)_{t=1}^{T}$ for $i=1,...,N$.
Here, $\sv_i^t=[\xv_i;\vv_i^t]$ is the \emph{representation} of user $i$ at time $t$, which is a sample in the state space $\mathcal{S}$ describing user preferences;
$\av_i^t$ is a vector of the estimated user preferences for user $i$, which is a sample in the action space $\mathcal{A}$ of the recommender.
Accordingly, we can model the recommendation-feedback loop as a Markov Decision Process (MDP) $\mathcal{M} = \langle\mathcal{S}, \mathcal{A}, P, R \rangle$, where ${P}:~\mathcal{S}\times\mathcal{A}\times\mathcal{S}\mapsto \mathbb{R}$ is the transition probability of user preferences and $R: \mathcal{S}\times\mathcal{A}\mapsto \mathbb{R}$ is the reward function used to evaluate recommended items.
We further assume that the recommender $\pi(\cdot)$ works as a policy parametrized by $\thetav$, $i.e.$, $\pi_\thetav(\av|\sv)$, which corresponds to the distribution of the estimated item preferences $\av$, conditioned on the user representation $\sv$. 
The target recommender should be an optimal policy that maximizes the expected reward: $J(\pi_\theta) = \sum_{t=1}^{T}\mathbb{E}_{\pi_\thetav}\left[R_\phiv(\sv^t, \av^t)\right]$,
where $R_\phiv(\sv^t, \av^t)$ means the reward for the state-action pair $(\sv^t, \av^t)$.
For the $i$-th user, given $\sv_i^t$, the recommender selects potentially-preferred items by finding the optimal item-preference vector as follows
\begin{eqnarray}\label{Eq:predictor}
\begin{aligned}
\av_i^t = \arg\sideset{}{_{\av\in\mathcal{A}}}\max \pi_\theta(\av|\sv_i^t).
\end{aligned}
\end{eqnarray}
and then recommending the top few items with largest values in the vector $\av_i^t$. Note that, different from traditional reinforcement learning tasks, in which both $\mathcal{S}$ and $\mathcal{A}$ are available while $P$ and $R$ are with limited accessibility, our recommender receives only \emph{partial} information of the state --- it does not observe users' feedback embedding $\vv_i$. 
In other words, to optimize the recommender, we need to build a reward model and a feedback generator \emph{jointly}, which motivates us to introduce a virtual user into the framework.

\subsection{The virtual user}
The virtual user module aims to implement the feedback function $f(\cdot)$ in Eq.~(\ref{eq:recsys}), which not only models the reward of the items provided by the recommender but also generates feedback $\vv_i^t$ to complete the representations of the state $\sv_i^t=[\xv_i;\vv_i^t]$.
Accordingly, the virtual user contains the following two modules:

\noindent\textbf{Reward Estimator} The reward estimator parametrizes the function of reward, which takes the current estimation $\av_i^t$ and user preference $\sv_i^t$ as input and evaluate their compatibility. 
In this work, we assume that the reward estimator is parametrized by parameters $\phiv$, and is defined as
\begin{eqnarray}\label{Eq:critic_score}
\begin{aligned}
R_{\phiv}(\sv_i^t, \av_i^t) = \text{sigmoid}(g(h(\xv_i, \av_i^t))).
\end{aligned}
\end{eqnarray}
For the reward estimator, we use the static part of the state $\sv_i^t$, $i.e.$, the observed user preference $\xv_i$ as input. 
In Eq.~(\ref{Eq:critic_score}), $h(\cdot,\cdot)$ denotes the fusion function which merges $\xv_i$ and $\av_i^t$ into a real value vector (the fused input is shown in Figure~\ref{fig:fused_function} and is described in the Appendix), and $g(\cdot)$ is the single value regression function that translates the fused input into a single reward value. 
The sigmoid function squashes the predicted reward value between 0 and 1.

\noindent\textbf{Feedback Generator}
The feedback generator connects the reward estimator with the recommender module via generating a feedback embedding, $i.e.$,
\begin{eqnarray}\label{Eq:feedbackgen}
\begin{aligned}
\vv^i_{t+1} = F_{\psiv}(h(\xv_i, \av_{i}^t), R_\phi(\sv_i^t, \av_i^t)),
\end{aligned}
\end{eqnarray}
where $\psiv$ represents the parameters of the generator.
Specifically, the parametric function $F_{\psiv}(\cdot,\cdot)$ considers the fused input and the estimated reward and returns a feedback embedding $\vv^i_t\in R^d$ to the recommender. In this work, we use a multilayer perceptron to model the function $F_{\psiv}$. In Eq.~(\ref{Eq:feedbackgen}),
$R_\phi(\sv_i^t, \av^i_t)$ is as the scalar reward (as defined in Eq.~(\ref{Eq:critic_score})) denoting the compatibility between the recommended items and user preferences, and $h(\xv_i, \av^i_t)$, which is a vector rather than a scalar like reward, further enriches the information of the reward to generate feedback embeddings.
Consequently, the recommender receives the informative feedback as a complementary component of the static observation $\xv_i$ to make an improved recommendation via Eq.~(\ref{Eq:predictor}).

\section{The Learning Algorithm}
\subsection{Learning task}
\begin{figure*}[t!]
	\centering
	\includegraphics[width=\textwidth]{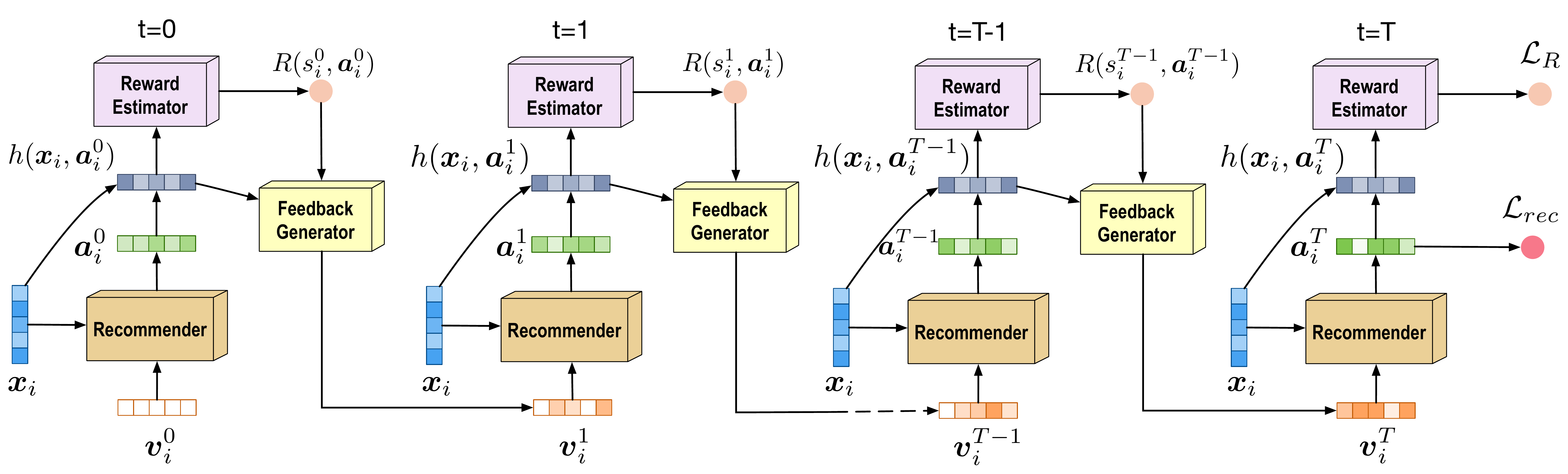}
	\vspace{-2em}
	\caption{\small{Unrolling the recurrent CF-SFL framework into an iterative learning process with  $T$ time steps.}}
	\label{fig:CAC_unrolled}
\end{figure*}
Based on the proposed framework, we need to jointly learn the policy corresponding to the recommender $\pi_\thetav$, the reward estimator $R_\phiv$, and the feedback generator $F_\psiv$.
Assuming we have a set of user observations $\mathcal{D}=\{\xv_i\}$, where $\xv_i\in \mathbb{R}^{M}$ is the vector of observed user preferences for user $i$. 
We formulate the learning task as the following min-max optimization problem
\begin{eqnarray}\label{eq:opt}
\begin{aligned}
\sideset{}{_{\pi_\thetav,  F_\psiv}}\min \sideset{}{_{R_\phiv}}\max
 \mathcal{L}(\pi_\thetav, R_\phiv, F_\psiv),
\end{aligned}
\end{eqnarray}
where
\begin{eqnarray}\label{eq:obj}
{
		\begin{aligned}
		&\mathcal{L}(\pi_\thetav, R_\phiv, F_\psiv)=\underbrace{\sideset{}{_{i}}\sum\mathcal{L}_{\text{rec}} (\av_i, \xv_i; \pi_\thetav,  F_\psiv)}_{\text{Reconstruction loss}} ~\nonumber \\ &-\underbrace{\mathbb{E}_{\av\sim \pi_\theta} [\log(R_\phiv(\sv,\av))]  -\mathbb{E}_{\av\sim \mathcal{D}} [1-\log(R_\phiv(\sv,\av))]}_{\text{Collaboration with adversarial regularizer}}
		\end{aligned} }
\end{eqnarray}
In particular, the first term $\mathcal{L}_{\text{rec}}$ in Eq.~(\ref{eq:obj}) can be any reconstruction loss based on user preferences $\mathcal{D}$, $e.g.$, the evidence lower bound (ELBO) proposed in VAEs~\cite{liang2018variational} (and used in our work). 
This term ensures the recommender to fit the observed user preference.
The second term considers the following types of interactions among the various modules:
\begin{itemize}
	\vspace{-1mm}
	\item The \emph{collaboration} between the recommender policy $\pi_\thetav$ and the feedback generator $F_\psiv$ towards a better predictive recommender. 
	\vspace{-1mm}
	\item The \emph{adversarial game} between the recommender policy $\pi_\thetav$ and the reward estimator $R_\phiv$.
	\vspace{-1mm}
\end{itemize}
Accordingly, given the current reward model, we update the recommender policy $\pi_\thetav$ and the feedback generator $F_\psiv$ to maximize the expected reward derived from the generated user representation $\sv$ and the estimated user preference $\av$. Likewise, given the recommended policy and the feedback generator, we improve the reward estimator $R_{\phiv}$ by sharpening its criterion --- the updated reward estimator maximizes the expected reward derived from the generated user representation and the observed user preference, while minimizing the expected reward based on the generated user representation and the estimated user preference. 

Therefore, we solve Eq.~(\ref{eq:opt}) via alternating optimization, updating of $\pi_\thetav$ and $F_\psiv$ by minimizing
\begin{eqnarray}\label{eq:loss_rec_fbgenrator}
\begin{aligned}
\mathcal{L}_C(\pi_\thetav, F_\psiv) = &\sideset{}{_{i}}\sum\mathcal{L}_{\text{rec}} (\av_i, \xv_i; \pi_\thetav,  F_\psiv) \\ &-\mathbb{E}_{\av\sim \pi_\theta} [\log(R_\phiv(\sv,\av))].
\end{aligned}
\end{eqnarray}
We update $R_{\phiv}$ is achieved by maximizing
\begin{eqnarray}\label{eq:loss_rewarder}
\begin{aligned}
\mathcal{L}_A(R_\phiv) = &-\mathbb{E}_{\av\sim \pi_\theta} [\log(R_\phiv(\sv,\av))] \\ &-\mathbb{E}_{\av\sim \mathcal{D}} [1-\log(R_\phiv(\sv,\av))].
\end{aligned}
\end{eqnarray}
Both these update steps can be solved efficiently via stochastic gradient descent.

\subsection{Unrolling for learning and inference}
Because the proposed framework contains a closed loop among learnable modules, during training we unroll the loop and let the recommender interact with the virtual user in $T$ steps.
Specifically, at the initial stage, the recommender takes the observed user preference vector $\xv_i$ and an all-zero initial feedback embedding vector $\vv_i^0$, to make recommendations. 
At each step $t$, the recommender outputs the estimated user preferences $\av^{t}_i$ given $\xv_i$ and $\vv^t_i$ to the virtual user, and receives the feedback embedding $\vv^{t+1}_i$.
The loss is defined according to the output of the last step, $i.e.$, $\av^T$ and $\vv^T$, and the modules are updated accordingly. 
After the model is learned, in the testing phase we need to infer the recommended item in the same manner, unrolling the feedback loop and deriving $\av^T$ as the final estimated user preferences. 
The details of unrolling process are illustrated in Figure~\ref{fig:CAC_unrolled}, and the detailed scheme of our learning algorithm is shown in Algorithm~\ref{alg:CF-SFL} in the Appendix. 

\section{CF-SFL as Inverse Reinforcement Learning}
Our CF-SFL framework automatically discovers informative user feedback as side information and gradually improve the training for the recommender. Theoretically, it is closely related to Inverse Reinforcement Learning (IRL). 
Specifically, we jointly learn the reward estimator $R_\phi$ and the policy (recommender) $\pi_\thetav$ from the {\em expert trajectories} $\mathcal{D}$ (\textit{i.e.,} the observed labeled data), which typically consists of state-action pairs generated from some expert policy $\pi_E$ with the corresponding environment dynamics $\rho_E$.
The IRL is to recover the optimal reward function $R^*$ as well as the optimal recommender $\pi^*$. Formally, the IRL is defined as:
\begin{eqnarray}
\begin{aligned}
\{R^*, \pi^*\} &\triangleq \text{IRL}(\pi_E) = 
		\arg\max_{\phiv} \sum_{\sv, \av}\rho_{E}(\sv, \av)R_\phiv(\sv, \av)\\ &-[\max_{\thetav}H(\pi) + \sideset{}{_{\sv, \av}}\sum\rho(\sv, \av)R_\phiv(\sv, \av)],
\end{aligned}
\end{eqnarray}
which can be rewritten as
\begin{eqnarray*}
\begin{aligned}
\max_{\phiv} \min_{\thetav} 
		\underbrace{\sideset{}{_{\sv, \av}}\sum(\rho_{E}(\sv, \av) - \rho(\sv, \av))R_\phiv(\sv, \av) -H(\pi)}_{\mathcal{L}(\pi, R)}
\end{aligned}
\end{eqnarray*}

Intuitively, the objective enforces the expert policy $\pi_E$ to induce higher rewards (the $\max$ part), than all other policies. This objective is sub-optimal if the expert trajectories are noisy, \textit{i.e.}, the expert is not perfect and its trajectories are not optimal. This will make the learned policy always perform worse than the expert one. Besides, the illed-defined IRL objective often induces multiple solutions due to the flexible solution space, \textit{i.e.}, one can assign an arbitrary reward to trajectories not from expert, as long as these trajectories yield lower rewards than the expert trajectories. To alleviate these issues, some constraints can be incorporated into the objective functions, {\it e.g.}, a convex reward functional, $\psi: \mathbb{R}^{\mathcal{S}\times\mathcal{A}} \rightarrow \mathbb{R}$, which usually works as a regularizer. 
\begin{align}\label{eq:IRL_max_min}
\{R^*, \pi^*\} = \arg\sideset{}{_{\phiv}}\max \sideset{}{_{\thetav}}\min\mathcal{L}(\pi_\thetav, R_\phiv)  - \psi(R_\phiv).
\end{align}

To imitate the expert policy and provide better generalization, we adopt the adversarial regularizer~\cite{ho2016generative}, which defines $\psi$ with the following form:
\begin{align*}
\psi(R_\phiv) \triangleq \left\{
\begin{array}{ll}
\mathbb{E}_{\pi_E}\left[q(R_\phiv(\sv, \av))\right] & \mbox{if } R_\phiv(\sv, \av) \geq 0\\
+\infty & \mbox{otherwise}
\end{array}~,
\right.
\end{align*}
where $q(x) = x - \log(1 - e^{-x})$. 
This regularizer places low penalty on reward functions $R$ that assign an amount of positive value to expert state-action pairs; however, if $R$ assigns low value (close to zero, which is the lower bound) to the expert, then the regularizer will heavily penalize $R_\phiv$. 
With the adversarial regularizer, we obtain a new imitation learning algorithm for the recommender:
\begin{align}
\sideset{}{_{\theta}}\min \psi^*(\rho_\pi-\rho_{\pi_E}) - \lambda H(\pi_\theta)
\end{align}
Intuitively, we want to find a saddle point $(R_{\phi},\pi_\theta)$ of the expression:
\begin{eqnarray*}
\begin{aligned}
\mathbb{E}_{\pi_\theta} [\log(R_\phiv(\sv,\av))] + \mathbb{E}_{\pi_E} [1-\log(R_\phiv(\sv,\av))] - \lambda H(\pi_\theta),
\end{aligned}
\end{eqnarray*}
where $R_\phiv(\sv,\av)\in(0,1)$. Note that Eq.~(\ref{eq:IRL_max_min}) is derived from the objective of traditional IRL. However, distinct from the traditional approach, we propose a feedback generator to provide feedback to the recommender. 
In terms of the reward estimator, it tends to assign lower rewards to the predicted results by the recommender $\pi_\theta$ and higher rewards for the expert policy $\pi_E$, which aims to discriminate $\pi_\theta$ from $\pi_E$, similar to Eq.~(\ref{eq:loss_rewarder}):
\begin{eqnarray}\label{Eq: Critic}
\begin{aligned}
\mathcal{L}_R = \mathbb{E}_{\pi_\theta} [\log(R_\phiv(\sv,\av))] + \mathbb{E}_{\pi_E} [1-\log(R_\phiv(\sv,\av))].
\end{aligned}
\end{eqnarray}

Similar to standard IRL, we update the generator to maximize the expected reward with respect to $\log R_\phiv(\sv,\av)$, moving towards expert-like regions of user-item space. 
In practice, we incorporate feedback embedding to update the user preferences,  and  the objective of the recommender is:
\begin{align}
\mathcal{L}_F &= \mathbb{E}_{\pi_\theta} [-\log(R([\xv_i, \vv_i^t],\av))] - \lambda H(\pi_\theta)
~\label{Eq:collaboration2}
\end{align}
where $\vv_i^t = F_\psi(h(\xv_i, \av_{i}^t), R_\phiv(\sv_i^t, \av_i^t))$. It is obvious that $\mathcal{L}_F$ recovers the second term in Eq.~(\ref{eq:loss_rec_fbgenrator}).

\begin{table}
\small
	\centering
	\caption{
		Dataset statistics 
	}\label{tab:dataset}
	\vspace{-1em}
	\begin{threeparttable}[c]
		\begin{tabular}{
				@{\hspace{4pt}}c@{\hspace{4pt}}|
				@{\hspace{4pt}}c@{\hspace{4pt}}
				@{\hspace{4pt}}c@{\hspace{4pt}}
				@{\hspace{4pt}}c@{\hspace{4pt}}
				@{\hspace{4pt}}c@{\hspace{4pt}}
				@{\hspace{4pt}}c@{\hspace{4pt}}
			}
			\hline\hline
			&\textbf{ML-20M} &\textbf{Netflix} & \textbf{MSD} \\ \hline
			$\#$ of users           & 136,677 & 463,435 & 571,355 \\ 
			$\#$ of items           &  20,108 & 17,769  & 41,140  \\
			$\#$ of interactions    &  10.0M  & 56.9M   & 33.6M   \\
			$\#$ of held-out-users  &  10.0K  & 40.0K   &  50.0K  \\
			$\%$ of sparsity        &  0.36$\%$ & $0.69\%$  & $0.14\%$ \\ 
			\hline\hline
		\end{tabular}
	\end{threeparttable}
	\vspace{-1.5em}
\end{table}

\section{Related Work}
\textbf{Collaborative Filtering (CF). } 
Existing CF approaches primary operate in one of the following two settings: CF with implicit feedback~\cite{bayer2017generic,hu2008collaborative} and CF with explicit feedback~\cite{koren2008factorization,liu2010unifying}. 
In implicit CF, user-item interactions are binary in nature (\textit{i.e}., 1 if clicked and 0 otherwise) as opposed to explicit CF where they represent item ratings (e.g., 1-5 stars). 
The implicit CF setting is more common/natural in many applications and has been widely studied, examples including factorization of user-item interactions~\cite{he2016fast,koren2008factorization,liu2016learning,rendle2010factorization,rennie2005fast} and ranking based approach~\cite{rendle2009bpr}. 
Our CF-SFL framework is also designed for the implicit CF. 

Currently, neural network based models have achieved state-of-the-art performance for various recommender systems~\cite{cheng2016wide,he2018outer,he2017neural,zhang2018neurec,liang2018variational}. Among these methods, NCF~\cite{he2017neural} casts the matrix factorization algorithm into an entire neural framework, combing the shallow inner-product based learner with a series of stacked nonlinear transformations. This method outperforms various traditional baselines and has motivated many follow-up works such as NFM~\cite{he2017neural}, Deep FM~\cite{guo2017deepfm} and Wide and Deep~\cite{cheng2016wide}. 
Recently, deep generative has also achieved remarkable success. In particular, VAE based approach to CF uses variational inference to scale up the algorithm for large-scale dataset and has shown significant success in recommender systems using multinomial~\cite{liang2018variational} or negative binomial~\cite{zhao2020variational} likelihoods. 
CF-SFL is a general framework which can be integrated with such models seamlessly.

\begin{table}
	\centering
	\caption{
		Architecture of our CF-SFL framework.  
	}\label{tab:architecture}
	\vspace{-1em}
	\begin{threeparttable}[c]
		\begin{small}
			\begin{tabular}{
					@{\hspace{3pt}}c@{\hspace{3pt}} |
					@{\hspace{3pt}}c@{\hspace{3pt}} |
					@{\hspace{3pt}}c@{\hspace{3pt}} 
					@{\hspace{3pt}}c@{\hspace{3pt}}
				}
				\hline\hline
				\textbf{Recommender} & \textbf{Reward Est.} & \textbf{Feedback Gen.} \\
				Input $\mathcal{R}^M$       & Input $\mathcal{R}^{64}$  &  Input $\mathcal{R}^{65}$ \\ \hline 
				$M\times 600$, tanh     & $64\times 128$, ReLU  & $64\times 128$, ReLU \\
				$600\times 200$ (x2)    & $128\times 128$, ReLU  & \\
				Sample $\mathcal{R}^{200}$  &                           & $128\times 128$, ReLU \\
				$200\times 600$, tanh   & $128\times 128$, ReLU &  \\
				$600\times M$ softmax       & $128\times 1$, sigmoid&  $128\times 128$\\
				\hline\hline
			\end{tabular}
		\end{small}
	\end{threeparttable}
\end{table}

\textbf{RL in CF. }
For RL-based methods,  
contextual multi-armed bandits have been utilized to model the interactive nature of recommender systems.  Thompson Sampling (TS)~\cite{chapelle2011empirical,kveton2015cascading,zhang2017learning} and Upper Confident Bound (UCB)~\cite{li2010contextual} are used to balance the trade-off between exploration and exploitation. Matrix factorization is combined with a bandit set-up in \cite{zhao2013interactive} to include latent vectors of items and users for better exploration. The MDP-Based CF model can be viewed as a partial observable MDP (POMDP) with partial observation of user preferences \cite{sunehag2015deep}. Value function approximation and policy based optimization can be employed to solve the MDP.  Modeling web page recommendation as a Q-Learning problem was proposed in \cite{zheng2018drn} and \cite{taghipour2008hybrid} and to make recommendations from web usage data.  An agent based approach was introduced in \cite{sunehag2015deep}  to address sequential decision problems. A novel page-wise recommendation framework based on deep reinforcement learning was proposed in \cite{zhao2018recommendations}. 
In this paper, we consider the recommending procedure as sequential interactions between virtual users and recommender; and leverage feedback from virtual users to improve the recommendation. 

\begin{table*}[t]
	\centering
	\caption{
		Performance comparison between our CF-SFL framework and various baselines. 
		$\text{VAE}^{*}$ is the results based on our own runs and $\text{VAE}^{\dagger}$ is the VAE model with our reward estimator.
	}\label{table:results_gain}
	\vspace{-1em}
		\begin{tabular}{
				@{\hspace{4pt}}l@{\hspace{4pt}} |
				c@{\hspace{4pt}}c@{\hspace{4pt}}c|
				c@{\hspace{4pt}}c@{\hspace{4pt}}c|
				c@{\hspace{4pt}}c@{\hspace{4pt}}c
			}
			\hline\hline
			\multirow{2}{*}{Methods} & \multicolumn{3}{c@{\hspace{4pt}}|}{ML-20M}& \multicolumn{3}{c@{\hspace{4pt}}|}{Netflix}& \multicolumn{3}{c}{MSD} \\
			&R@20  &R@50  &NDCG@100  
			&R@20  &R@50  &NDCG@100 
			&R@20  &R@50  &NDCG@100 \\
			\hline
			SLIM        & 0.370& 0.495& 0.401  & 0.347 & 0.428 & 0.379  &- & - & -  \\
			WMF  & 0.360& 0.498& 0.386  & 0.316 & 0.404 & 0.351  & 0.211 & 0.312 & 0.257   \\
			CDAE & 0.391& 0.523& 0.418  & 0.343 & 0.428 & 0.376  & 0.188 & 0.283 & 0.237   \\
			aWAE & 0.391 & 0.532 & 0.424 & 0.354 & 0.441 & 0.381 & - & - & - \\
			VAE & 0.395& 0.537& 0.426  & 0.351 & 0.444 & 0.386  & 0.266 & 0.364 & 0.316  \\ \hline
			$\text{VAE}^{*}$ & 0.395& 0.535& 0.425   & 0.350 & 0.444 & 0.386 & 0.260 & 0.356 & 0.311 \\ 
			$\text{VAE}^{\dagger}$ & 0.396 & 0.536 & 0.426  & 0.352 &0.445 &0.387   & 0.263 & 0.360 & 0.314 \\ 
			CF-SFL &\textbf{0.404} &\textbf{0.542} & \textbf{0.435}   &\textbf{0.355} &\textbf{0.449} & \textbf{0.394}   &\textbf{0.273} &\textbf{0.369} &\textbf{0.323}   \\ 
			\hline\hline
		\end{tabular}
\end{table*}

\begin{figure*}[t]
	\centering
	\includegraphics[width=\linewidth, height=130pt]{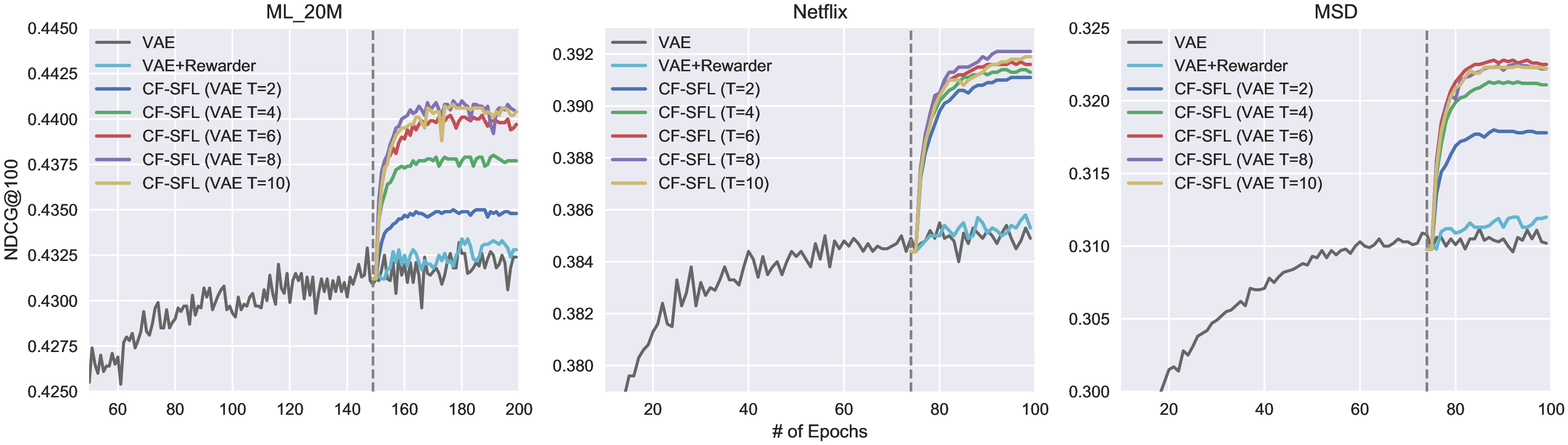}
	\vspace{-1em}
	\caption{\small{Performance (NDCG@100) boost on the validation sets.}} 
	\label{fig:performanceBoost}
\end{figure*}

\section{Experiments}
\textbf{Datasets}
We investigate the effectiveness of the proposed CF-SFL framework on three benchmark datasets of recommendation systems. 
(\textit{i}) MovieLens-20M (ML-20M), taken from a movie recommendation service containing tens of millions user-movie ratings; 
(\textit{ii}) Netflix-Prize (Netflix), another user-movie ratings dataset collected by the Netflix Prize~\cite{bennett2007netflix}; 
(\textit{iii}) Million Song Dataset (MSD), a user-song rating dataset, which is released as part of the Million Song Dataset~\cite{bertin2011million}. 
To directly compare with existing work, we employed the same pre-processing procedure as~\cite{liang2018variational}. 
A summary statistics of these datasets are provided in Table~\ref{tab:dataset}. 

\begin{figure*}[t]
	\centering
	\includegraphics[width=\linewidth, height=140pt]{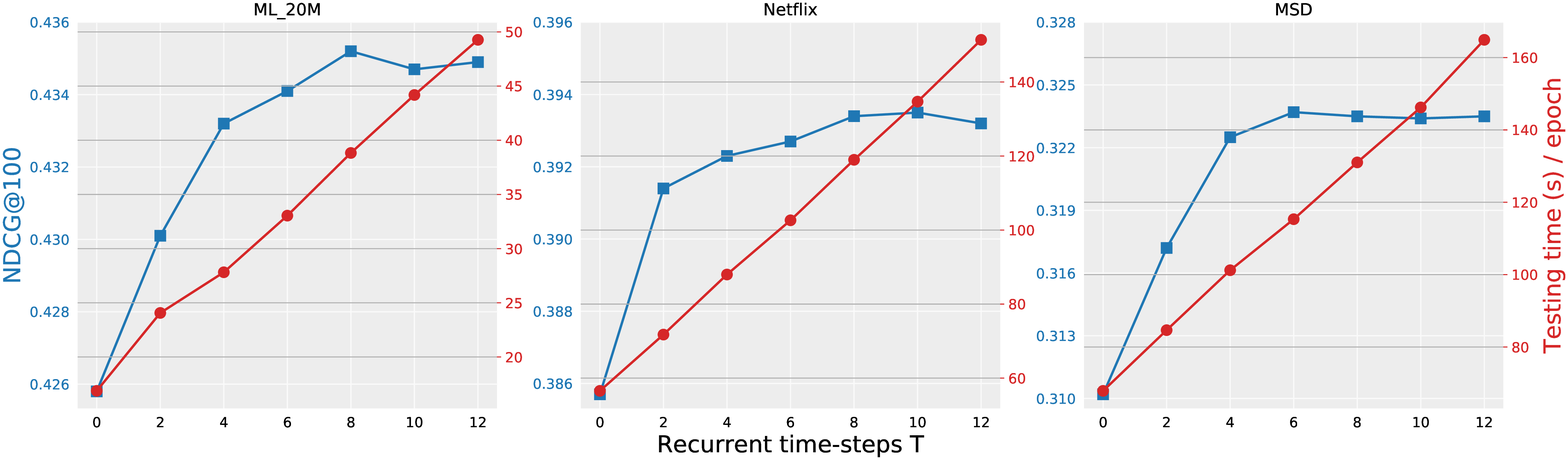}
	\caption{ \small{The blue curve summarizes NDCG@100 and red curves report the computational cost for model inference in each epoch. In each sub-figure, we vary the time steps from 0 to 12 ($T=0$ is the base recommender).} }
	\label{fig:varyT}
\end{figure*}

\textbf{Evaluation Metrics}
We employ Recall@r\footnote{\url{https://en.wikipedia.org/wiki/Precision_and_recall}} together with NDCG@r\footnote{\url{https://en.wikipedia.org/wiki/Discounted_cumulative_gain}} as the evaluation metric for recommendation, which measures the similarity between the recommended items and the ground truth. 
Recall@r considers top-r recommended items equally, while NDCG@r ranks the top-r items and emphasizes the importance of the items that are with high ranks.

\textbf{Set-up}
For our CF-SFL framework, the architectures of its recommender, reward estimator and feedback generator are shown in Table~\ref{tab:architecture}.
To represent the user preference, we normalize $\xv_i$ and $\vv_i^t$ ($t>0$) independently and concatenate the two into a single vector. 
To learn the model, we pre-train the recommender (150 epochs for ML-20M and 75 epochs for Netflix and MSD) and optimize the entire framework (50 epochs for ML-20M and 25 epochs for the other two). 
$\ell_2$ regularization with a penalty term $0.01$ is applied to the recommender, and Adam optimizer~\cite{kingma2014adam} with batch in size of $500$ is employed. 

\textbf{Baselines}
To demonstrate the efficacy of our framework, we consider multiple state-of-the-art approaches as baselines, which can be categorized into two types:
(\textit{i}) Linear models: SLIM~\cite{ning2011slim} and WMF~\cite{hu2008collaborative}; and
(\textit{ii}) Deep neural network based models: CDAE~\cite{wu2016collaborative}, VAE~\cite{liang2018variational}, and aWAE~\cite{zhong2018wasserstein}. 
It should be noted that our CF-SFL is a generalized framework, which is compatible with all these approaches. 
In particular, as shown in Table~\ref{tab:architecture}, we implement our recommender as the VAE-based model~\cite{liang2018variational} for a fair comparison. 
In the following experiments, we will show that besides such a setting the recommender can be implemented by other existing models as well.

All the evaluation metrics are averaged across all the test sets.

\textit{(i)} \textbf{Quantitative Results:}
we test various methods and report their results in Table~\ref{table:results_gain}. 
With the proposed CF-SFL framework, we observe improvements over the baselines on all the evaluation metrics. 
These experimental results demonstrate the power of the proposed CF-SFL framework, which provides informative feedback as side information. 
Particularly, we observed that the performance of the base model (VAE$^*$) is similar to that of its variation with the reward estimator (VAE$^\dagger$). 
It implies that simply learning a feedback from the reward estimator via back-propagation is not much helpful. 
Compared with such a na\"ive strategy, the proposed CF-SFL provides more informative feedback to the recommender, and is able to  improve recommendation results more effectively.

\textit{(ii)} \textbf{Learning Comparison:}
In Figure~\ref{fig:performanceBoost}, we show the training  trajectory of the baselines (VAE, VAE+reward estimator) and the CF-SFL with multiple time steps. 
There are several interesting findings: 
(a) The performance of the base VAE doesn't improve after the pre-training steps, \textit{e.g.,} 75 epochs for Netflix. In comparison, the proposed CF-SFL framework can further improve the performance once the whole model is triggered; 
(b) The CF-SFL yields fast convergence once the whole framework is activated; (c) Consistent with the results in Table~\ref{table:results_gain}, the trajectory of VAE$^\dagger$ in Figure~\ref{fig:performanceBoost} is similar to that of the base VAEs (VAE$^*$). 
In contrast, the trajectories of our CF-SFL methods are more smooth and are to converge to a much better local minimum. 
This phenomenon further verifies that our CF-SFL  learns informative user feedback with better stability; (d) With an increase in the number of time steps $T$ in a particular range ($T \leq 8$ for ML-20M), CF-SFL achieves faster and better performance; 
One possible explanation is the learning with our unrolled structure --- parameters are shared across different time-steps, and a more accurate gradient is found towards the local minimum; and
(e) We find \text{ML-20M} and \text{MSD} are more sensitive to the choice of $T$ when compared with Netflix. Therefore, the choice of $T$ should be adjusted for different datasets. 

\textit{(iii)} \textbf{CF-SFL with Dynamic Time Steps:}
As shown in Figure~\ref{fig:CAC_unrolled}, learning of CF-SFL involves a recurrent structure with $T$ times steps. 
We investigate the choice of $T$ and report its influence on the performance of our method.
Specifically, the NDCG$@$100 with different $T$ values is shown in Figure~\ref{fig:varyT}. 
Within 6 time steps, CF-SFL consistently boots the performance on all the three datasets. 
Even with a larger time step, the results remain stable. 
Additionally, the inference time of CF-SFL is linear in $T$. 
To achieve a trade-off between performance and efficiency, in our experiments we set $T$ to $8$ for ML-20M and Netflix and $6$ for MSD. 

\begin{table}
\small
	\centering
	\caption{
		Comparisons for various recommenders.
	}\label{tab:other_actors}
	\vspace{-1em}
	\begin{threeparttable}[c]
		\begin{tabular}{
				@{\hspace{2pt}}c@{\hspace{2pt}}|
				@{\hspace{2pt}}c@{\hspace{2pt}}
				@{\hspace{2pt}}c@{\hspace{2pt}}
				@{\hspace{2pt}}c@{\hspace{2pt}}
				@{\hspace{2pt}}c@{\hspace{2pt}}
				@{\hspace{2pt}}c@{\hspace{2pt}}
			}
			\hline\hline
			\textbf{Recommender} &\textbf{w/o CF-SFL} & \textbf{w CF-SFL} & \textbf{Gain ($10^{-3}$)}\\ \hline
			WARP  &0.31228 & 0.33987 & +27.59\\
			MF &0.41587 & 0.41902 & +3.15 \\
			DAE&0.42056 & 0.42307 & +2.51 \\ \hline
			VAE&0.42546 & 0.43472 & +9.26\\
			VAE-(Gaussian)   &0.42019 & 0.42751 & +7.32\\
			VAE-($\beta=0$)&0.42027 & 0.42539 & +5.02\\
			VAE-Linear     &0.41563 & 0.41597 & +0.34\\ 
			\hline\hline
		\end{tabular}
	\end{threeparttable}
	\vspace{-1em}
\end{table}

\textbf{Relative Improvements due to Generative Feedback}
As mentioned earlier, our CF-SFL is a generalized framework which is compatible with many existing collaborative filtering approaches. 
We study the usefulness of our CF-SFL on various recommendation systems and present the results in Table~\ref{tab:other_actors}.
Specifically, two types of recommenders are being considered: linear approaches like WARP~\cite{weston2011wsabie} and MF~\cite{hu2008collaborative}, and deep learning methods, \textit{e.g.}, DAE~\cite{liang2018variational} and the variation of VAE in~\cite{liang2018variational}. 
We find that our CF-SFL is capable of generalizing most such collaborative filtering approaches and boosts their performance accordingly. 
The gains achieved by our CF-SFL may vary depending on the choice of recommender. 

\section{Conclusion}
We presented CF-SLF, a novel framework for making recommendation from sparse data by simulating user feedback.
It constructs a virtual user to provide informative side information as user feedback. 
We formulate the framework as an IRL problem and learn the optimal policy by feeding back the action and reward. 
Specifically, a recurrent architecture was built to unrolled the framework for efficient learning. 
Empirically we improve the performance of state-of-the-art collaborative filtering methods with a non-trivial margin.
Our framework serves as a practical solution making IRL feasible over large-scale collaborative filtering. 
It will be interesting to investigate the framework in other applications, such as sequential recommender systems.

\section{Acknowledgment} 
The Duke University component of this work was supported in part by DARPA, DOE, NIH, ONR and NSF, and a portion of the work performed by the first two authors was performed when they were affiliated with Duke. 

\clearpage
{
\bibliography{reference.bib}
}

\clearpage
\appendix
\section{Appendix}
\begin{algorithm}[ht]
	\SetNoFillComment
	\caption{CF-SFL training with stochastic optimization}
	\label{alg:CF-SFL}
	\begin{algorithmic}[1]
		\STATE \textbf{Input:} A user-item matrix $\Xv$ and observed data $\mathcal{D}$, the  unrolling step $T$, the size of batch $b$. 
		\STATE \textbf{Output:} Recommender $\pi_\thetav$, reward estimator $R_\phiv$, and feedback generator $F_\psiv$
		\STATE \textbf{Initialization}: randomly initialize $\thetav$, $\phiv$ and $\psiv$; \\
		\tcc{\scriptsize{stage 1: pretrain the recommender}}
		\WHILE{not converge}
		\STATE Sample a batch of $\{\xv_i \}_{i=1}^b$ from $\mathcal{D}$;
		\STATE Update $\thetav$ via minimizing $\mathcal{L}_{rec}$.
		\ENDWHILE
		\BlankLine \tcc{\scriptsize{stage 2: pretrain the reward estimator}}
		\WHILE{not converge}
		\STATE Sample a batch of $\{\xv_i \}_{i=1}^b$ from $\mathcal{D}$ and calculate  $\{R_\phiv(\sv_i, \av_{i})\}_{i=1}^b$;
		\STATE Sample another batch of user $\{\xv_i\}_{i=1}^b$ and set $\vv_i=\bm{0}$
		\STATE Infer the recommended items $\{\av_i\}_{i=1}^b$ and calculate $\{R_\phiv(\sv_i, \av_{i})\}_{i=1}^b$;
		\STATE Update $\phiv$ via maximizing (\ref{eq:loss_rewarder}).
		\ENDWHILE
		\BlankLine \tcc{\scriptsize{stage 3: alternative train all the modules}}
		\WHILE{not converge}
		\STATE Sample a batch of $\{\xv_i \}_{i=1}^b$ from $\mathcal{D}$, initialize feedback embedding $\vv^{0} = \textbf{0}$;
		\BlankLine \tcc{\scriptsize{Update recommender and feedback generator}}
		\STATE Feed $\{\xv_i\}_{i=1}^b$ and $\Vv^{(0)}$ into the recommender and infer $\{\av_i^{T}\}_{i=1}^b$ through a $T$-step recurrent structure. 
		\STATE Collect the corresponding reward $\{R_\phiv(\sv_i^T, \av_i^T)\}_{i=1}^b$
		\STATE Update $\thetav$ and $\psiv$ via minimizing (\ref{eq:loss_rec_fbgenrator}).
		\BlankLine \tcc{\scriptsize{Reward estimator update step}}
		\STATE Sample a batch of $\{\xv_i \}_{i=1}^b$ from $\mathcal{D}$, and calculate $\pi(\xv_i)$ and $\{R_\phiv(\sv_i, \av_i \}_{i=1}^b$;
		\STATE Sample a batch of $\{\xv_i\}_{i=1}^b$, infer the recommended items $\{\av_i^T\}_{i=1}^b$ and calculate $\{R_\phiv(\sv_i, \av_{i}^T)\}_{i=1}^b$;
		\STATE Update $\phiv$ via maximizing (\ref{eq:loss_rewarder})
		\ENDWHILE
	\end{algorithmic}
\end{algorithm}

\subsection{Fusion function}~\label{Sec:fusedfunction}
Here we give a detail description of the fusion function we have proposed. A straightforward way to build the fusion function $h(\xv_i, \av_i)$ is to concatenate $\xv_i$ and $\av_i$, and feed it into a linear layer to learn a lower dimensional representation. 
However, in practice, this method is infeasible since the dimension of items, $M$, is extremely large and using concatenation will make the problem even worse. To this end, we introduce a sparse layer. This layer includes a lookup table $B\in \mathcal{R}^{M\times d}$. Once we have inferred the recommended items $\av_i$ based on the observation $\xv_i$, we build the the fused input as 
\begin{eqnarray}
\begin{aligned}
h(\xv_i, \av_i) = \frac{1}{|\xv_i|}\sideset{}{_{j=1}^M}\sum \delta(\xv_{ij}) B_j + \sideset{}{_{k=1}^M}\sum \av_{ik} B_k
\end{aligned}
\end{eqnarray}

\begin{figure}
  \begin{center}
    \includegraphics[width=\linewidth]{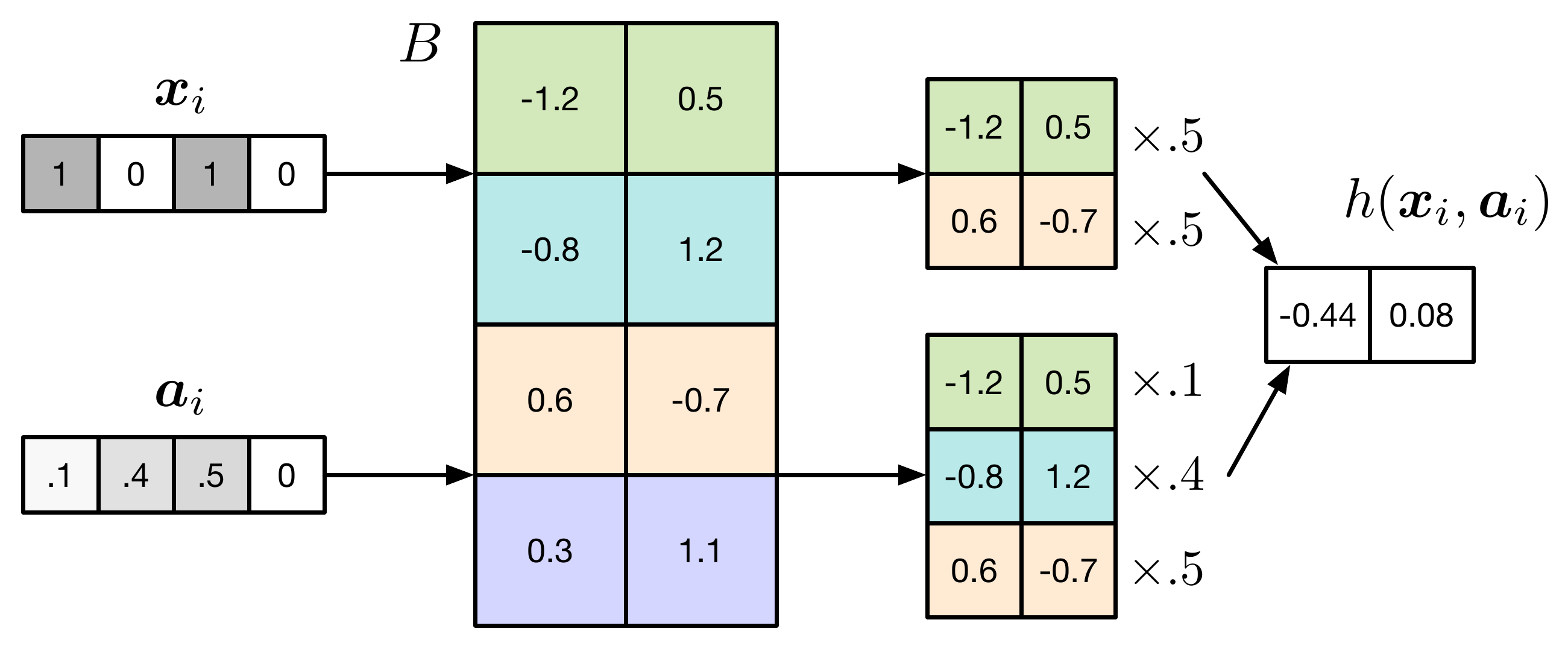}
    \vspace{-1em}
  \end{center}
  \caption{\small An example of the our fused function working scheme. The user preference $\xv_i$ and the recommended items $\av_i$ share the same lookup table $B$. $[-0.04, 0.08]$ is the fused input for the given example. This method works efficient if $\xv_i$ and $\av_i$ are sparse. }\label{fig:fused_function}
\end{figure}
where $\delta$ is the Dirac Delta function and takes value 1 if $\xv_{ij}=1$, $|\xv_i|$ is number of 1 in $\xv_i$. The parameters of the lookup table will be automatically learned during the training phrase. We show an example to illustrate the working scheme for the proposed fusion function in Figure~\ref{fig:fused_function}. The benefits for the proposed approach are as follows: 1) it reduces the computational cost of the standard linear transformation under the general sparse set up and saves number of parameters in our proposed adversarial learning framework; 2) This lookup table is shared across the observations and the recommended items, building a unified space for the users' existing preferences and missing preferences. Empirically such shared knowledge boosts the performance of our CF-SFL framework.

\end{document}